\documentclass[9pt,twocolumn,twoside]{osajnl}

\journal{ol} 

\setboolean{shortarticle}{true}

\newcommand\wone{\hat{\mathbb{I}}}
\newcommand\wG{\hat{\mathbb G}}
\newcommand\wM{\hat{\mathbb M}}
\newcommand\wP{\hat{\mathbb P}}
\newcommand\wQ{\hat{\mathbb Q}}
\newcommand\wD{\hat{\mathbb D}}
\newcommand\wS{\vec{\mathbb S}}
\newcommand\wF{\vec{\mathbb F}}
\newcommand\wJ{\vec{\mathbb J}}

\newcommand\one{\hat{\mathbf{1}}}
\newcommand{\be}{\begin{equation}}
\newcommand{\ee}{\end{equation}}
\newcommand{\bea}{\begin{eqnarray}}
\newcommand{\eea}{\end{eqnarray}}
\newcommand{\Eq}[1]{Eq.\,(\ref{#1})}

\newcommand{\br}{\mathbf{r}}
\newcommand{\bF}{\mathbf{F}}
\newcommand{\bE}{\mathbf{E}}
\newcommand{\bH}{\mathbf{H}}
\newcommand{\bD}{\mathbf{D}}
\newcommand{\bB}{\mathbf{B}}
\newcommand{\bj}{\mathbf{j}}
\newcommand{\bJ}{\mathbf{J}}

\newcommand{\heps}{\hat{{\pmb{\varepsilon}}}}
\newcommand{\hmu}{\hat{{\pmb{\mu}}}}
\newcommand{\hxi}{\hat{{\pmb{\xi}}}}
\newcommand{\hzeta}{\hat{{\pmb{\zeta}}}}
\newcommand{\heta}{\hat{{\pmb{\eta}}}}

\title{Resonant-state expansion for open optical systems: Generalization to magnetic, chiral, and  bi-anisotropic materials}

\author[1,*]{E.\,A. Muljarov}
\author[2]{T. Weiss}

\affil[1]{School of Physics and Astronomy, Cardiff University, Cardiff CF24 3AA,
United Kingdom}
\affil[2]{4th Physics Institute and Research Center SCoPE,
University of Stuttgart, Pfaffenwaldring 57, D-70550 Stuttgart, Germany}

\affil[*]{Corresponding author: Egor.Muljarov@astro.cf.ac.uk}

\ociscodes{(140.4780) Optical resonators; (030.4070) Modes; (160.1585) Chiral media; (160.3918) Metamaterials.}

\begin{abstract}
The resonant-state expansion, a recently developed powerful method in electrodynamics, is generalized here for open optical systems containing magnetic, chiral, or bi-anisotropic materials. It is shown that the key matrix eigenvalue equation of the method remains the same, but the matrix elements of the perturbation now contain variations of the permittivity, permeability, and bi-anisotropy tensors. A general normalization of resonant states in terms of the electric and magnetic fields is presented.
\end{abstract}

\setboolean{displaycopyright}{true}

\begin{document}

\maketitle

The resonant-state expansion (RSE) is a novel powerful theoretical method that has been recently developed in electrodynamics~\cite{MuljarovEPL10}.
The RSE is a rigorous perturbation theory, not limited to small perturbations, which warrants an efficient calculation of all resonant states (RSs) of an open optical system in an arbitrarily selected spectral range. This calculation is based on knowing the RSs of another, so-called basis system, which is usually (but not necessarily) simpler than the system of interest, ideally having an exact analytic solution.  The RSE was verified and tested on optical systems of different shape and dimensionality~\cite{DoostPRA12,DoostPRA13,DoostPRA14,ArmitagePRA14,LobanovPRA17}, demonstrating its superior computational efficiency~\cite{DoostPRA14,LobanovPRA17} compared to available numerical methods, such as finite-difference time-domain~\cite{YeeIEEE66,Kunz93}, finite element~\cite{Dhatt12}, and aperiodic Fourier modal method~\cite{LalanneOL00,PisarencoJOSAA10}.

Being originally introduced in nuclear physics almost a century ago~\cite{GamowZP1928,SiegertPR39}, RSs in electrodynamics present eigensolutions of Maxwell's equations satisfying outgoing boundary conditions, which correspond to electromagnetic excitations decaying in time, with the electromagnetic energy leaking out of the system. This leakage, however, causes an exponential growth of the RS wave function with distance, so that the standard normalization, used e.g. for bound states in quantum mechanics or for waveguide modes in optics, diverges.
While the correct normalization for scalar fields was known~\cite{Shnol1971}, expressions for the normalization of the electromagnetic vector fields of the RSs, intensively used in the literature~\cite{LeungJOSAB96,KristensenOL12}, are only approximate, as has been recently clarified~\cite{MuljarovPRB16a,MuljarovPRA17}. The correct normalization of RSs in finite optical systems, which is a cornerstone of the RSE, was presented in the very first paper on the method~\cite{MuljarovEPL10} and was later generalized to arbitrary systems with frequency dispersion of the permittivity~\cite{MuljarovPRB16}. Recently, it has been used to formulate an exact theory of the Purcell effect~\cite{MuljarovPRB16a}. Furthermore, the exact normalization was extended  to photonic crystal structures~\cite{WeissPRL16,WeissPRB17} and applied to resonantly enhanced refractive index sensing using the RSE with only one and two RSs in the basis, providing surprisingly accurate descriptions.

The RSE has been also generalized to optical systems with frequency dispersion of the permittivity~\cite{MuljarovPRB16} without affecting the computational complexity, which is a very important step towards describing realistic materials and specifically plasmonic effects. This was achieved by treating the dispersion as an analytical function with a finite number of simple poles in the lower half-plane of the complex frequency, known in the literature as generalized Drude-Lorentz model~\cite{SehmiPRB17}.

So far, the RSE was applied to non-magnetic optical systems, fully described by its permittivity. Naturally, the RS normalization and the RSE itself dealing with perturbations of the permittivity were formulated in terms of the electric field only.  However, the most general materials with local response are bi-anistropic and have non-zero magnetic susceptibility and coupling tensors between electric and magnetic fields, including the chiral optical activity and circular dichroism~\cite{Lindell1994}.  Describing such systems, which include but are not limited to metamaterials~\cite{KrieglerIEEE10}, chiral plasmonics~\cite{GovorovChem12,NesterovACS16} and chiral sensors~\cite{HentscheleSA17} is of growing interest.
It is therefore crucial to have a general formulation of the RSE and the RS normalization, in which the electric and magnetic fields contribute as equal partners, and the local linear response of an optical system is taken in the most general form. This is done in the present work below.
\medskip

\noindent {\large\bf Maxwell's equations and Green's dyadic.} An arbitrary linear optical system is described by Maxwell's equations in a medium:
\be
\nabla \times\bE= i k \bB\,,
\ \ \ \
\nabla \times\bH= -ik \bD+\frac{4\pi}{c}\bj\,,
\label{ME1}
\ee
where $k=\omega/c$ is the wave number in vacuum, and $\omega$ is the frequency of the electromagnetic field.
Quite generally, for systems with a spatially local linear response, one can write
\be
\bD= \heps \bE+\hxi \bH\,,
\ \ \ \
 \bB= \hmu \bH+\hzeta \bE\,,
\label{Blin}
\ee
with frequency dependent tensors of permittivity $\heps(k,\br)$ and permeability $\hmu(k,\br)$, and bi-anisotropy tensors $\hxi(k,\br)$ and $\hzeta(k,\br)$.
In the following, we concentrate on  systems satisfying the reciprocity relation, leading additionally to $\heps^{\rm T}=\heps$, $\hmu^{\rm T}=\hmu$, and $\hxi^{\rm T}=-\hzeta$, where T denotes tensor transposition.
Equations~(\ref{ME1}--\ref{Blin}) can be written in the following compact symmetric way:
\be
\wM(k,\br) \wF(\br)=\wJ(\br)\,,
\label{ME}
\ee
where  $\wM(k,\br)=k\wP(k,\br)-\wD(\br)$ is  a 6x6 matrix operator, with
\be
\wP(k,\br)=\left(\begin{array}{cc}
\heps&\heta\\
\heta^{\rm T}&\hmu\\
\end{array}\right)\,,
\ \ \ \
\wD(\br)= \left(\begin{array}{cc}
0&\nabla\times\\
\nabla\times&0\\
\end{array}\right)\,,
\ee
and $\heta=-i\hxi$. The electric and magnetic fields as well as the currents are now represented by 6-dimensional vectors,
\be
\wF(\br)= \left(\begin{array}{c}
\bE\\
i\bH\\
\end{array}\right)
\ \ \ {\rm and}\ \ \
\wJ(\br)= \left(\begin{array}{c}
 \bJ_E\\
i\bJ_H
\end{array}\right)\,,
\ee
respectively, where $\bJ_E=-4\pi i\bj/c$, and the magnetic current $\bJ_H$ is introduced for symmetry purposes (although this is not necessary).

We now introduce a generalized dyadic Green's function (GF) $ \wG_k(\br,\br')$
with outgoing boundary conditions in the regions outside the optical system, satisfying the equation
\be
\wM(k,\br) \wG_k(\br,\br')=\wone\delta(\br-\br')\,,
\label{GF-equ}
\ee
in which $\wone$ is the 6x6 identity matrix.
The GF has simple poles~\cite{DoostPRA13,DoostPRA14} at $k=k_n$, which are the wave numbers of the RSs of the system. The RSs are in turn the eigen solutions of the homogeneous Maxwell's equations,
\be
\wM(k_n,\br) \wF_n(\br)=0\,,
\label{RS-equ}
\ee
satisfying outgoing boundary conditions, where the index $n$ is used to label the RSs. Owing to the reciprocity principle and the Mittag-Leffler (ML) theorem, the GF is represented as a series~\cite{DoostPRA13}
\be
\wG_k(\br,\br')=\sum_n \frac{\wF_n(\br)\otimes \wF_n(\br')}{k-k_n}\,,
\label{GF-ML}
\ee
determinining the normalization of RSs that is considered below.
Note that \Eq{GF-ML} is valid within the system, or rather within a minimal convex volume including it.
\medskip

\noindent  {\large\bf Closure relation and sum rules.} Substituting the ML expansion \Eq{GF-ML} into \Eq{GF-equ} for the GF and using \Eq{RS-equ}, we obtain
\be
\sum_n \frac{k\wP(k,\br)-k_n\wP(k_n,\br)}{k-k_n}  \wF_n(\br)\otimes \wF_n(\br')=\wone\delta(\br-\br')\,.
\label{closure1}
\ee
In the absence of dispersion, 
\Eq{closure1} immediately results in the following closure relation:
\be
\wP(\br)\sum_n \wF_n(\br)\otimes \wF_n(\br')=\wone\delta(\br-\br')\,.
\label{closure2}
\ee
In the case of a frequency dispersion described by a generalized Drude-Lorentz model~\cite{MuljarovPRB16,SehmiPRB17},
the matrix $\wP$ becomes
\be
\wP(k,\br)=\wP_\infty(\br) +\sum_j \frac{\wQ_j(\br)}{k-\Omega_j}\,,
\label{DL}
\ee
having complex poles at $k=\Omega_j$ with generalized conductivities $\wQ_j(\br)$.
Substituting \Eq{DL} into \Eq{closure1} and
using the algebraic identity
\be
\frac{1}{k-k_n}\left(\frac{k}{k-\Omega_j}-\frac{k_n}{k_n-\Omega_j}\right)= \frac{-\Omega_j}{(k-\Omega_j)(k_n-\Omega_j)}\,,
\label{algebra1}
\ee
yields
\be
\!\sum_n\!\!\left[\wP_\infty(\br) -\sum_j \frac{\Omega_j \wQ_j(\br)}{(k-\Omega_j)(k_n-\Omega_j)} \right]\!\!\wF_n(\br)\otimes \wF_n(\br')
=\wone\delta(\br-\br').
\label{closure3}
\ee
Since the Lorentzian functions are linearly independent, \Eq{closure3} splits into sum rules
\be
\wQ_j(\br)\sum_n \frac{\wF_n(\br)\otimes \wF_n(\br')}{k_n-\Omega_j}=0
\label{sum}
\ee
and a closure relation
\be
\wP_\infty(\br)\sum_n \wF_n(\br)\otimes \wF_n(\br')=\wone\delta(\br-\br')\,,
\label{closure4}
\ee
similar to the non-dispersive one, \Eq{closure2}. Summing  \Eq{sum} over all $j$ and adding it to \Eq{closure4}, we can reformulate the closure relation as
\be
\sum_n \wP(k_n,\br) \wF_n(\br)\otimes \wF_n(\br')=\wone\delta(\br-\br')\,.
\label{closure5}
\ee
\medskip

\noindent  {\large\bf Normalization of resonant states.} As already mentioned, the form of the GF \Eq{GF-ML} determines the normalization of the RS wave functions $\wF_n(\br)$. To derive this normalization, we introduce an analytic continuation $\wF(k,\br)$ of RS field $\wF_n(\br)$ in the complex $k$-plane around the $k=k_n$ point. $\wF(k,\br)$ satisfies \Eq{ME}, which can be solved with the help of the GF. More specifically, using \Eq{GF-ML}, we obtain
\be
\wF(k,\br)=\int  \wG_k(\br,\br') \wJ(\br') d\br'
= \sum_n\frac{\wF_n(\br)}{k-k_n} \int  \wF_n(\br') \cdot \wJ(\br') d\br'\,.
\ee
The requirement that $\wF(k,\br)\to \wF_n(\br)$  in the limit $k\to k_n$ results in the following $k$ dependence of the current: $\wJ(\br)= (k-k_n) \wS(\br)$, where $\wS(\br)$  can be chosen as a $k$-independent function, normalized in such a way that
\be
\int  \wF_n(\br) \cdot \wS(\br) d\br=1\,.
\label{S-norm}
\ee
Equation~(\ref{S-norm}) then provides the normalization of the RSs. Indeed, multiplying \Eq{ME} with $\wF_n(\br)$, \Eq{RS-equ} with $\wF(k,\br)$, and taking the difference between the two, yields
\bea
&&k\wF_n\cdot \wP(k)\wF-k_n\wF\cdot \wP(k_n)\wF_n
\nonumber\\
&&-\wF_n\cdot \wD(k)\wF+\wF\cdot \wD(k_n)\wF_n=(k-k_n)\wF_n\cdot \wS\,,
\label{S-norm1}
\eea
where the $k$ and $\br$ dependencies are dropped for brevity of notations.
The third and the fourth terms in the left hand side of \Eq{S-norm1} can be written as
$$
-\wF_n\cdot \wD(k)\wF+\wF\cdot \wD(k_n)\wF_n=i\nabla\cdot(\bE_n\times \bH-\bE\times \bH_n)\,.
$$
Integrating \Eq{S-norm1} over an arbitrary volume $V$ containing the system, using the divergence theorem, and taking the limit $k\to k_n$, we obtain a general formula for the RS normalization:
\be
1=\int_V \wF_n\cdot [k \wP(k)]'\wF_n d\br+i\oint_{S_V} \left(\bE_n\times \bH'_n-\bE'_n\times \bH_n\right)\cdot d{\bf S}\,,
\label{Norm1}
\ee
where $S_V$ is the boundary of $V$, and the prime means the derivative with respect to $k$ taken at $k=k_n$. Differentiation of the matrix $k \wP(k)$ is straightforward, whereas the derivatives of the analytic continuation of the fields outside the system can be expressed as~\cite{MuljarovEPL10,MuljarovPRB16a}
\be
\bF'_n=\frac{1}{k_n} (\br\cdot\nabla) \bF_n\,,
\label{Fprime}
\ee
in which $\bF_n$ is either $\bE_n$ or $\bH_n$. The normalization \Eq{Norm1} then takes an explicit form in terms of the electric and magnetic fields of a given RS:
\bea
1&=&\int_V \left[ \bE_n\cdot (k \heps)'\bE_n+ \bE_n\cdot (k \hxi)'\bH_n\right] d\br
\nonumber\\
&&-\int_V \left[ \bH_n\cdot (k \hzeta)'\bE_n+ \bH_n\cdot (k \hmu)'\bH_n\right] d\br
\label{Norm2}\\
&&+\frac{i}{k_n}\oint_{S_V}\left[\bE_n\times (\br\cdot\nabla)\bH_n+\bH_n\times (\br\cdot\nabla)\bE_n\right]\cdot d{\bf S}\,.
\nonumber
\eea
This general normalization is fully consistent with the analytic normalizations we have previously used in terms of the electric field~\cite{MuljarovEPL10,DoostPRA14,MuljarovPRB16a,MuljarovPRB16}, for systems described by the permittivity, as we demonstrate below.
We note however that writing the GF as in \Eq{GF-ML}, the electric field of the normalized RS is a factor of $\sqrt{2}$ smaller than the one used in our previous works.
Furthermore, as we also show below, the general normalization \Eq{Norm2} is suited for both static and non-static RSs, which is consistent with two different expression used previously for these cases~\cite{DoostPRA14}.
\medskip

\noindent  {\large \bf Normalization of RSs in terms of the electric field.} Let us show that for non-magnetic materials, described by only the permittivity, the general normalization \Eq{Norm2} reduces to the one previously used in terms of the electric field only~\cite{MuljarovEPL10,DoostPRA14,MuljarovPRB16a,MuljarovPRB16}. In this case $\hxi=\hzeta=0$ and $\hmu=\one$, where $\one$ is a 3x3 identity matrix, and \Eq{Norm2} becomes
\bea
1&=&\int_V \bE_n\cdot (k \heps)'\bE_n\, d\br - \int_V \bH_n\cdot \bH_n\,d\br
\nonumber\\
&&+i\oint_{S_V}\left(\bE_n\times \bH'_n-\bE'_n\times \bH_n\right)\cdot d{\bf S}\,,
\label{Norm3}
\eea
where we have taken the surface term again in the form of the field derivatives, as in \Eq{Norm1}.
Using the Poynting theorem for the RS wavefunction, we can transform the second volume integral in \Eq{Norm3} into
\be
- \int_V  \bH_n\cdot \bH_n\,d\br= \frac{i}{k_n}  \oint_{S_V} \bE_n\times \bH_n \cdot d{\bf S}+\int_V \bE_n \cdot \heps\bE_n\,d\br\,.
\ee
For the surface integral in \Eq{Norm3} we obtain
\bea
&&i\oint_{S_V}\left(\bE_n\times \bH'_n-\bE'_n\times \bH_n\right)\cdot d{\bf S}
=-\frac{i}{k_n}  \oint_{S_V} \bE_n\times \bH_n \cdot d{\bf S}
\nonumber\\
&&+\frac{1}{k_n}\oint_{S_V}\left(\frac{\partial \bE'_n}{\partial s}\cdot \bE_n-\frac{ \partial \bE_n}{\partial s}\cdot \bE'_n\right) dS\,,
\label{interm2}
\eea
using vector identities, as well as $\nabla \times\bE'_n= i \bH_n+i k_n \bH'_n$ and
the fact that $\nabla\cdot\bE_n= \nabla\cdot\bE'_n = 0$ outside the system.
Collecting all terms, we obtain the normalization condition for RSs with $k_n\neq0$:
\be
1=2\int_V \bE_n\cdot \left.\frac{\partial (k^2\heps)}{\partial (k^2)}\right|_{k_n}\!\!\bE_n\,d\br
+\frac{1}{k_n}\oint_{S_V}\!\left(\frac{\partial \bE'_n}{\partial s}\cdot \bE_n-\frac{ \partial \bE_n}{\partial s}\cdot \bE'_n\right) dS,
\label{Norm4}
\ee
where $\partial/{\partial s}$ means the spatial derivative along the surface normal, and  $\bE'_n =(\br\cdot\nabla) \bE_n/k_n$, according to \Eq{Fprime}.

For static electric modes with $k_n=0$, the condition $\bH_n=0$ leads to the volume term in \Eq{Norm3} with the magnetic field vanishing. Since the electric field of a static mode  $\bE_n\to 0$ far away from the system~\cite{DoostPRA14} and the surface of integration can be chosen as any closed surface including the system, one can get rid of the surface integral, ending up with the volume integral of the electric field over the entire space:
\be
1=\int  \bE_n\cdot \left.\frac{\partial (k^2\heps)}{\partial (k^2)}\right|_{k_n}\bE_n\,d\br\,.
\label{Norm-static}
\ee
Both results \Eq{Norm4} and \Eq{Norm-static} are identical to the normalization of resonant states in non-magnetic materials obtained in~\cite{MuljarovEPL10,DoostPRA14,MuljarovPRB16a,MuljarovPRB16}, with the already noted factor of 2 introduced in the present work.
\medskip

\noindent  {\large\bf Resonant-state expansion.}  Let us now consider a perturbed system described by a general frequency dependent perturbation $\Delta\wP(k,\br)$ of the permittivity, permeability, and bi-anisotropy tensors. The Maxwell equation for a perturbed RS $\wF(\br)$ then takes the form:
\be
\wM(k,\br) \wF(\br)=-k\Delta\wP(k,\br) \wF(\br)\,,
\label{ME-pert}
\ee
where $k$ is the perturbed eigenvalue. Note that the unperturbed system and the perturbation are chosen in such a way that the perturbation is included in the minimal convex volume containing the unperturbed system. Solving \Eq{ME-pert} with the help of the GF, we obtain
\be
\wF(\br)=-k\int \wG_k(\br,\br') \Delta\wP(k,\br') \wF(\br') d\br'\,.
\label{Dyson1}
\ee
Let us first assume a non-dispersive perturbation $\Delta\wP(\br)$.
Substituting the ML expansion \Eq{GF-ML} into \Eq{Dyson1} and expanding the perturbed field inside the system into the unperturbed RSs as
\be
\wF(\br)= \sum_n c_n \wF_n(\br)\,,
\ee
we obtain
\be
\sum_n c_n \wF_n(\br)= -k \sum_n\frac{\wF_n(\br)}{k-k_n}\sum_m V_{nm} c_m\,,
\label{interm}
\ee
where the matrix elements of the perturbation are given by
\be
V_{nm}=\int \wF_n(\br)\cdot \Delta\wP(\br)\wF_m(\br)d\br\,.
\ee
Equating coefficients at the basis functions $\wF_n(\br)$, \Eq{interm} reduces to
\be
(k-k_n) c_n = -k \sum_m V_{nm} c_m \,,
\label{RSE1}
\ee
which is the standard non-dispersive RSE equation~\cite{MuljarovEPL10,DoostPRA14}.

Taking into account the dispersion of the perturbation in a generalized Drude-Lorentz form,
\be
\Delta\wP(k,\br)= \Delta\wP_\infty(\br)+\sum_j \frac{\Delta\wQ_j(\br)}{k-\Omega_j},
\label{D-L}
\ee
\Eq{Dyson1} becomes
\bea
\wF(\br)&=&-k\int \wG_k(\br,\br') \Delta\wP_\infty(\br') \wF(\br') d\br'\,,
\nonumber\\
&&-k\sum_j \int \wG^j_k(\br,\br') \frac{\Delta\wQ_j(\br')}{k-\Omega_j} \wF(\br') d\br'\,,
\label{Dyson2}
\eea
where we have added in the second line zeros in the form of the sum rules defined by \Eq{sum}:
\be
\wG^j_k(\br,\br') =\wG_k(\br,\br') + \frac{\Omega_j}{k}
\sum_n \frac{\wF_n(\br)\otimes \wF_n(\br')}{k_n-\Omega_j}\,.
\label{Gj}
\ee
Using again the ML expansion \Eq{GF-ML} of the GF $\wG_k(\br,\br')$ and the algebraic identity~\Eq{algebra1},
we arrive, after equating coefficients at the basis functions $\wF_n(\br)$,
at the linear eigenvalue equation of the dispersive RSE:
\be
(k-k_n) c_n = -k \sum_m V_{nm}(\infty) c_m
+k_n \sum_m [V_{nm}(\infty)-V_{nm}(k_n)] c_m
\label{RSE2}
\ee
with the matrix elements of the dispersive perturbation given by
\be
V_{nm}(k)=\int \wF_n(\br)\cdot \Delta\wP(k,\br)\wF_m(\br)d\br\,.
\label{Vnm}
\ee
Note that \Eq{RSE2} has exactly the same form as that developed in~\cite{MuljarovPRB16}, and in case of no frequency dispersion it reduces back to \Eq{RSE1}.
However, the matrix elements \Eq{Vnm} now have the most general form, which can be written explicitly as
\bea
V_{nm}(k)&=&\int_V \left[ \bE_n\cdot \Delta\heps(k)\bE_m+ \bE_n\cdot \Delta\hxi(k)\bH_m\right] d\br
\label{Vnm2}\\
&&-\int_V \left[ \bH_n\cdot \Delta\hzeta(k)\bE_m+ \bH_n\cdot \Delta\hmu(k) \bH_m\right] d\br\,.
\nonumber
\eea
The matrix elements \Eq{Vnm2} are expressed in terms of the electric and magnetic fields of basis RSs $n$ and $m$ and generally dispersive changes of the tensors of the permittivity $\Delta\heps(k,\br)$, permeability $\Delta\hmu(k,\br)$, and bi-anisotropy couplings $\Delta\hxi(k,\br)$ and  $\Delta\hzeta(k,\br)$ between the electric and magnetic fields.
Solving the matrix eigenvalue problem \Eq{RSE2} of the RSE determines the wave numbers $k$ of the perturbed RSs and the coefficients $c_n$ of the expansion of the perturbed wave functions into the known RSs of a basis system.  Presently, this is the most efficient and intuitive computational approach for finding the RSs of open optical systems, as demonstrated in numerous publications~\cite{MuljarovEPL10,DoostPRA12,DoostPRA13,DoostPRA14,ArmitagePRA14,LobanovPRA17,MuljarovPRB16,WeissPRL16,WeissPRB17}. This approach is now generalized to bi-anisotropic systems.
\smallskip

In conclusion, we have generalized the resonant-state expansion for open optical systems containing arbitrary reciprocal bi-anisotropic materials or metamaterials, including those having magnetic and chiral optical activity, as well as circular dichroism. We have presented the theory in the most general, compact and symmetrized way, with the electric and magnetic field vectors contributing on equal footing. We have addressed both cases of non-dispersive systems and systems having frequency dispersion described by a generalized Drude-Lorentz model. We have derived a general compact expression for the normalization of resonant states, expressed in terms of the electric and magnetic fields and shown its equivalence to the one used previously for systems fully described by their permittivity and expressed in terms of the electric field only. The presented theory has the widest spectrum of applications, ranging from modeling and optimization of chirality sensors to accurate description of the optics of magnetic and metamaterial systems.

\bigskip

\noindent
{\large\bf Acknowledgements.} E. A. M. acknowledges discussions with W. Langbein and support by the EPSRC Grant
EP/M020479/1, the S\^er Cymru National Research Network in Advanced Engineering and Materials, and RBRF Grant 16-29-03283. T. W. acknowledges support from DFG SPP 1839, VW Foundation, and the MWK Baden-W\"urttemberg.

\bibliography{genRSE}

\end{document}